\newread\testifexists
\def\GetIfExists #1 {\immediate\openin\testifexists=#1
    \ifeof\testifexists\immediate\closein\testifexists\else
    \immediate\closein\testifexists\input #1\fi}

\immediate\openin\testifexists=e
    \ifeof\testifexists\immediate\closein\testifexists\else
    \immediate\closein\testifexists\input e\fipsf

\magnification= \magstep1  
\tolerance=1600
\parskip=5pt
\baselineskip= 5 true mm \mathsurround=1pt
\font\smallrm=cmr8  
\font\medrm=cmr9  
\font\bigbf=cmbx12
    \def\Bbb#1{\setbox0=\hbox{$\tt #1$}  \copy0\kern-\wd0\kern .1em\copy0}
    \immediate\openin\testifexists=a
    \ifeof\testifexists\immediate\closein\testifexists\else
    \immediate\closein\testifexists\input a\fimssym.def 

\def\secbreak{\vskip12pt plus .6in \penalty-200\vskip -2pt plus -.4in}
\def\ref#1{${\,}^{\hbox{\smallrm #1}}$}
   \def\newsect#1{\secbreak\noindent{\bf #1}\medskip}
   
\def\hugeskip{\vskip12mm plus 3mm}
\def\Narrower{\par\narrower\noindent}   
\def\Endnarrower{\par\leftskip=0pt \rightskip=0pt}
    \def\ra{\rightarrow}

\def\cl{\centerline}    
\def\ni{\noindent}      \def\pa{\partial}   \def\dd{{\rm d}}
            \def\ket{\rangle}

\def\a{\alpha}      \def\b{\beta}         
\def\d{\delta}      \def\D{\Delta}  \def\e{\varepsilon}
               
\def\m{\mu}         \def\f{\phi}            \def\vv{\varphi}
\def\n{\nu}             
       
\def\t{\tau}          
              
\def\w{\omega}        

 \def\LL{{\cal L}}

\def\fn#1{\ifcase\noteno\def\fnchr{*}\or\def\fnchr{\dagger}\or\def
    \fnchr{\ddagger}\or\def\fnchr{\medrm\S}\or\def\fnchr{\|}\or\def
    \fnchr{\medrm\P}\fi\footnote{$^{\fnchr}$}
    {\scrunch#1\toe}\ifnum\noteno>4\global\advance\noteno by-6\fi
    \global\advance\noteno by 1}
    \def\scrunch{\baselineskip=11 pt \medrm}
    \def\toe{\vphantom{$p_\big($}}
    \newcount\noteno

\def\fract#1#2{{\textstyle{#1\over#2}}}
\def\ffract#1#2{\raise .35 em\hbox{$\scriptstyle#1$}\kern-.25em/
    \kern-.2em\lower .22 em\hbox{$\scriptstyle#2$}}

\def\half{\fract12}

\def\bbf#1{\setbox0=\hbox{$#1$} \kern-.025em\copy0\kern-\wd0
    \kern.05em\copy0\kern-\wd0 \kern-.025em\raise.0433em\box0}

\def\deff{\ {\buildrel{\rm def}\over{=}}\ }

{\ }\vglue 1truecm \rightline{SPIN-2001/07}
\rightline{hep-th/0104080} \hugeskip \cl{\bigbf DETERMINISM IN
FREE BOSONS} \hugeskip

\cl{Gerard 't Hooft }
\bigskip
\cl{Institute for Theoretical Physics} \cl{Utrecht University,
Leuvenlaan 4} \cl{ 3584 CC Utrecht, the Netherlands}
\smallskip
\cl{and}
\smallskip
\cl{Spinoza Institute} \cl{Postbox 80.195} \cl{3508 TD Utrecht,
the Netherlands}
\smallskip\cl{e-mail: \tt g.thooft@phys.uu.nl}
\cl{internet: \tt http://www.phys.uu.nl/\~{}thooft/} \hugeskip
\ni{\bf Abstract}\Narrower It is shown how to map the quantum
states of a system of free Bose particles one-to-one onto the
states of a completely deterministic model. It is a classical
field theory with a large (global) gauge group. \Endnarrower
\hugeskip
\newsect{1. Introduction.} Consider a model for free, relativistic
bosons, described by the Lagrangian
$$\LL=-\half(\pa_\m\f)^2-\half\m^2\f^2\,.\eqno(1.1)$$
We take this Lagrangian as our prototype, but Lorentz invariance
is not crucial; with slight modifications, our analysis will be
applicable just as well to non-relativistic free bosons. If we
compare it with a {\it classical\/} field theory, described by
the Klein-Gordon equation
    $$(\D-\m^2)\vv-\ddot\vv=0\,,\eqno(1.2)$$
where the dots refer to time differentiation, then we obviously
have a quite different physical system. One important difference
is that phase space for the classical system is, in a sense, twice
as big: at a given time $t=0$, one may specify the values of
$\vv(\vec x,0)$ and $\dot\vv(\vec x,0)$ independently, whereas the
quantized theory only requires us to specify the operators
$\f(\vec x,0)$.

The classical model is obviously invariant under the group of the
following transformations:
    $$\eqalign{\vv'(\vec x,t)&=\int\dd\vec
    y\,\big( K_1(\vec y)\,\vv(\vec x+\vec y,t)+K_2(\vec
    y)\,\dot\vv(\vec x+\vec y,t)\big)\,,\cr \dot\vv'(\vec
    x,t)&=\int\dd\vec y\,\big( K_1(\vec y)\,\dot\vv(\vec x+\vec
    y,t)+K_2(\vec y)\,(\D-\m^2)\vv(\vec x+\vec y,t)\big)\,,
    }\eqno(1.3)$$
where the integration kernels $K_1(\vec y)$ and $K_2(\vec y)$ are
arbitrary but fixed real generalized functions of $\vec y$,
independent of $\vec x$, and independent of $t$. They are
distributions, obeying certain integrability conditions. If
otherwise $K_1$ and $K_2$ were essentially arbitrary, then {\it
all\/} field configurations obeying the Klein-Gordon equation
could be transformed into any other solution. However, we impose
the following restrictions:
    $$K_1(\vec y)=K_1(-\vec y)\quad;\qquad K_2(\vec y)=-K_2(-\vec
    y)\,,\eqno(1.4)$$ In momentum space, the transformation then reads
    $$\eqalign{\hat\vv'(\vec k,t)&=\hat K_1(\vec k)\,\hat\vv(\vec k,t)
    +i\hat K_2(\vec k)\,\dot{\hat\vv}(\vec k,t)\,,\cr
    \dot{\hat\vv}'(\vec k,t)&=\hat K_1(\vec k)\,\dot{\hat\vv}(\vec
    k,t)-i\hat K_2(\vec k)\,({\vec k}^2+\m^2)\hat \vv(\vec k,t)\,,
    }\eqno(1.5)$$
where $\hat K_1(\vec k)$ is an even real function of $\vec k$, and
$\hat K_2(\vec k)$ an odd real function.

We now decide to call all functionals of $\vv(\vec x,t)$ that are
invariant under the transformation (1.3) with the constraint
(1.4) (which can easily be seen to form a group) `observables',
whereas all non-invariant quantities are fundamentally
unobservable. Notice that, because of the constraint (1.4), the
space of observables at $t=0$ is essentially half as big as the
classical phase space, just like the quantum theory. This is
easily seen in the following way:

In the original system, we were free to choose $\vv(\vec x,0)$
and its time derivative, $\dot\vv(\vec x,0)$, independently, after
which all field values at different times are fixed by the
Klein-Gordon equation. But now we may impose a `gauge condition',
such as
    $$\dot\vv(\vec x,0)=\d(\vec x)\quad\hbox{or}\qquad \dot{\hat\vv}(\vec k)
    =(2\pi)^{-3/2}\,.\eqno(1.6)$$
Inspecting the second line of Eq.~(1.5), we see that this
condition for $\dot{\hat\vv}'(\vec k,0)$ can be realized starting
from any set of values of  $\dot{\hat\vv}(\vec k,0)$ and
$\hat\vv(\vec k,0)$ by adjusting the real numbers $\hat K_1(\vec
k)$ and $\hat K_2(\vec k)$ in there, except for the singular cases
when the phases of $\dot{\hat\vv}(\vec k,0)$ and $\hat\vv(\vec
k,0)$ coincide, which happens only for a set of $\vec k$ values of
measure zero.

Hence this condition fixes\fn{Observe that this gauge choice is
not completely free from singularities and ambiguities, but those
are not relevant for the present discussion.} the kernels $K_1$
and $K_2$, so that the values of $\vv(\vec x,0)$ in this gauge all
correspond to observables.

Not only does the counting argument for the dimensionality of
phase space for this model match with the quantum mechanical case,
we claim an important theorem: \Narrower {\it The classical model
where the observables are restricted to be invariant under (1.3),
with constraints (1.4), is equivalent to the quantized
model.}\Endnarrower \ni We prove this theorem by first
considering a single harmonic oscillator.

If this theorem has any implications for hidden variable models,
we shall refrain from discussing that here\ref{1, 2}. It is just
the mathematical fact that we wish to expose.

\newsect{2. The harmonic oscillator revisited.}
The treatment of the harmonic oscillator to be given here, differs
somewhat from earlier treatments in this context\ref{2, 3},
although the philosophy is the same: we start with a deterministic
system whose evolution law is represented by a quantum
Hamiltonian.

The deterministic system we start with here is a set of $N$
states, $\{(0),(1),\cdots,(N-1)\}$ on a circle. Time is discrete,
the unit time steps having length $\t$ (the continuum limit is
left for later). The evolution law is:
    $$t\ra t+\t\quad:\qquad (\n)\ra(\n+1\,{\rm mod}\, N)\,.\eqno(2.1)$$
On the basis spanned by the states $(\n)$, the evolution operator
is
    $$U(\D t=\t)\ =\ e^{-iH\t}\ =e^{-\textstyle{\pi i\over
    N\vphantom{_g}}}\pmatrix{0\,&1\cr &0&1\cr
    &&\ddots&\ddots\cr&&&0&1\cr 1&&&&0\cr}\ .\eqno(2.2)$$
The phase factor in front of the matrix is of little importance;
it is there for future convenience. Its eigenstates are denoted as
$|n\ket$, $n=0,\cdots,N-1$. We have
    $$H|n\ket={2\pi(n+\half)\over N\t}|n\ket\,.\eqno(2.3)$$
The $\half$ comes from the aforementioned phase factor.

It is now instructive to apply the algebra of the $SU(2)$
generators $L_x$, $L_y$ and $L_z$, so we write
    $$N\deff 2\ell+1\quad,\qquad n\deff m+\ell\quad,\qquad
    m=-\ell,\cdots,\ell\ .\eqno(2.4)$$
Using the quantum numbers $m$ rather than $n$ to denote the
eigenstates, we have
    $$H|m\ket={2\pi(m+\ell+\half )\over (2\ell+1)\t}|m\ket\qquad\hbox{or}\qquad
    H={\textstyle{2\pi\over (2\ell+1)\t}}\,(L_z+\ell+\half)\ .
    \eqno(2.5)$$
This Hamiltonian resembles the harmonic oscillator Hamiltonian
when \hbox{$\w=2\pi/(2\ell+1)\t$}, except for the fact that there
is an upper bound for the energy. This upper bound disappears in
the continuum limit, if $\ell\ra\infty$, $\t\downarrow 0$. Using
$L_x$ and $L_y$, we can make the correspondence more explicit.
Write
    $$\eqalign{ L_\pm|m\ket&\deff\sqrt{\ell(\ell+1)-m(m\pm1)}|m\pm1\ket\
    ;\cr  L_\pm&\deff L_x\pm iL_y \quad;\qquad[L_i,L_j]=i\e_{ijk}L_k\
    ,}\eqno(2.6)$$
and define
    $$x\deff\a L_x\quad,\qquad p\deff\b L_y\quad;\qquad
    \a\deff\sqrt{\t\over\pi}\quad,\qquad\b\deff{-2\over
    2\ell+1}\sqrt{\pi\over\t} \ .\eqno(2.7)$$
The commutation rules are
    $$[x,p]=\a\b iL_z=i(1-{\t\over\pi}H)\,,\eqno(2.8)$$
and since
    $$L_x^2+L_y^2+L_z^2=\ell(\ell+1)\,,\eqno(2.9)$$ we have
    $$H=\half\w^2 x^2+\half p^2+ {\t\over2\pi}\left({\w^2\over 4}+
    H^2\right)\,. \eqno(2.10)$$
Now consider the continuum limit, $\t\downarrow 0$, with
$\w=2\pi/(2\ell+1)\t$ fixed, for those states for which the
energy stays limited. We see that the commutation rule (2.8) for
$x$ and $p$ becomes the conventional one, and the Hamiltonian
becomes that of the conventional harmonic oscillator. There are
no other states than the legal ones, and their energies are
bounded, as can be seen not only from (2.10) but rather from the
original definition (2.5). Note that, in the continuum limit,
both $x$ and $p$ become continuous operators.

The way in which these operators act on the `primordial' or
`ontological' states $(\n)$ of Eq.~(2.1) can be derived from
(2.6) and (2.7), if we realize that the states $|m\ket$ are just
the discrete Fourier transforms of the states $(\n)$. This way,
also the relation between the eigenstates of $x$ and $p$ and the
states $(\n)$ can be determined, but we will not dwell on these
details.

The  most important conclusion from this section is that there is
a close relationship between the quantum harmonic oscillator and
the classical particle moving along a circle. The period of the
oscillator is equal to the period of the trajectory along the
circle. We started our considerations by having time discrete, and
only a finite number of states. This is because the continuum
limit is a rather delicate one. One cannot directly start with the
continuum because then the Hamiltonian does not seem to be bounded
from below.

\newsect{3. Multiple harmonic oscillators: the free Bose field.}

Extending our procedure to a collection of many harmonic
oscillators appears to be easy. We just take an equal number of
particles moving on circles. To be accurate, we must take the
time quantum $\t$ equal for all circles. We then have an
automaton, hopping from one state to the next at the beat of a
clock. but how do we handle coupled oscillators?

If the oscillators are coupled harmonicly, the prescription is
easy: we diagonalize the Hamiltonian, and handle all normal modes
independently. Then, we take as many circles as there are normal
modes. But then the question remains: how do we recognize these
circles in a realistic setting? In this paper, we set as our aim
the understanding of the free Bose field, which is nothing but an
infinite collection of harmonic oscillators, in terms of a
deterministic model. The biggest challenge then is, how to arrive
at a model that has a unique circular orbit for every normal mode
of the quantum model.

Consider the Lagrangian (1.1), and the associated Klein-Gordon
equation (1.2). The independent normal modes are the Fourier
coefficients $\hat\vv(\vec k,t)$, with
    $$\vv(\vec x,t)\deff (2\pi)^{-3/2}\int\dd\vec k\,\hat\vv(\vec
    k,t)\,e^{i\vec k\cdot\vec x}\ .\eqno(3.1)$$
If $\vv(\vec x,t)$ is a classical field, then its Fourier modes
$\hat\vv(\vec k,t)$ are all classical oscillators. They are not
confined to the circle, but the real parts, $\Re(\hat\vv(\vec
k,t))$, and the imaginary parts, $\Im(\hat\vv(\vec k,t))$ of
every Fourier mode each move in a two-dimensional phase space. If
we want to reproduce the quantum system, we have to replace these
two-dimensional phase spaces by one-dimensional circles.

It turns out to be possible to extract the circular component of
these oscillations --- and to remove their amplitudes!~--- but a
certain amount of care is needed. We do not want to destroy
translation invariance. This is why it is not advised to start
from the real part, $\Re(\hat\vv(\vec k))$, and the imaginary
part, $\Im(\hat\vv(\vec k))$, separately. Rather, we note that,
at each $\vec k$, there are two oscillatory modes, a positive and
a negative frequency. Thus, in general,
    $$\eqalign{\hat\vv(\vec k,t)&=A(\vec k)\,e^{i\w t}+B(\vec
    k)\,e^{-i\w t}\,;\cr \dot{\hat\vv}(\vec k,t)&=i\w A(\vec
    k)\,e^{i\w t}-i\w B(\vec k)\,e^{-i\w t}\,,}\eqno(3.2)$$
where $\w=(\vec k^2+\m^2)^{1/2}$. The most essential point of this
paper now is that we have to replace the amplitudes $A$ and $B$ by
numbers of modulus one, keeping only the circular motions $e^{\pm
i\w t}$. A space translation would mix up the real and imaginary
parts of $\hat\vv(\vec k,t)$ and $\dot{\hat\vv}(\vec k,t)$, which
is why we use the decomposition (3.2), where a space translation
merely rotates the phases $e^{\pm i\w t}$.

We introduce the `gauge transformations'
    $$\eqalign{A(\vec k)&\ra  R_1(\vec k)A(\vec k)\ ,\cr
    B(\vec k)&\ra R_2(\vec k)B(\vec k)\ ,}\eqno(3.3)$$
where $R_1(\vec k)$ and $R_2(\vec k)$ are {\it real\/} functions
of $\vec k$. The {\it only\/} quantities invariant under these
two transformations are the phases of $A$ and $B$, which is what
we want. In terms of $\vv$ and $\dot\vv$, this transformation
reads:
    $$\eqalign{\vv&\ra \textstyle
    {R_1+R_2\over 2}\,\vv+\textstyle{R_1-R_2\over
    2i\w}\,\dot\vv\ ,\cr \dot\vv&\ra \textstyle{R_1+R_2\over
    2}\,\dot\vv+\textstyle{i\w(R_1-R_2)\over 2}\,\vv\ .}\eqno(3.4)$$
This is how we arrive at the transformation (1.5), with
    $$\eqalign{\hat K_1(\vec k)&=\textstyle{R_1+R_2\over 2}\ ,\cr
    \hat K_2(\vec k)&=\textstyle{R_2-R_1\over 2\w}\ .}\eqno(3.5)$$

\newsect{4. Discussion.}
If we define `observables' to be all functionals of the fields
that are invariant under the transformations (1.3) with
restrictions (1.4), then, as was derived in Sect.~3, at each
Fourier mode, only the phase factors $e^{\pm i\w t}$ are
observable. These circular motions are the continuum limit of
discrete circular motions, and the latter span the Hilbert space
of  states that are described by the $SU(2)$ algebra of operators
$L_x$, $L_y$ and $L_z$. In the continuum limit, their Hamiltonian
is that of the harmonic oscillator, and these harmonic
oscillators combine into the system of quantized free Bose
particles. In contrast to our previous constructions, there is no
unwanted negative component of the Hamiltonian. It has been
successfully projected out by our invariance requirement.

Extending our procedure to systems with multiple bosons, or
vector bosons, may appear to be straightforward, except that the
gauge group may become difficult to reconcile with continuous
symmetries, such as rotation symmetry in a vector theory.
Therefore, vector theories are still posing a challenge.
Introducing interactions of any kind is an even bigger challenge.

\newsect{References.}

\item{1.} G.~'t~Hooft, ``Quantum Gravity as a Dissipative Deterministic System",
SPIN-1999/07, gr-qc/9903084; {\it Class.~Quant.~Grav. \bf 16}
(1999) 3263;

\item{2.} G.~'t~Hooft, ``Determinism and Dissipation in Quantum Gravity",
presented at {\it Basics and Highlights in Fundamental Physics},
Erice, August 1999, SPIN-2000/07, hep-th/0003005.

\item{3.} M.~Blasone, P.~Jizba and G.~Vitiello, ``Dissipation and
Quantization", hep-th/0007138.

 \bye